\documentstyle[12pt]{article}
\newcommand{\be}{\begin{eqnarray}}
\newcommand{\ee}{\end{eqnarray}}
\newcommand{\non}{\nonumber}

\begin{document}

\begin{titlepage}
\strut\hfill UMTG--198
\vspace{.5in}
\begin{center}

\LARGE Factorization of Multiparticle Scattering in the Heisenberg 
Spin Chain \\[1.0in]
\large Anastasia Doikou, Luca Mezincescu and Rafael I. Nepomechie\\[0.8in]
\large Physics Department, P.O. Box 248046, University of Miami\\[0.2in]
\large Coral Gables, FL 33124 USA\\

\end{center}

\vspace{.5in}

\begin{abstract}
We give an explicit proof within the framework of the Bethe 
Ansatz/string hypothesis of the factorization of multiparticle 
scattering in the antiferromagnetic spin-${1 \over 2}$ Heisenberg spin 
chain, for the case of 3 particles.
\end{abstract}

\end{titlepage}

\section{Introduction}

For models in $1+1$ dimensions, there are general arguments \cite{ZZ1} that 
integrability (i.e., the existence of nontrivial local integrals of 
motion) implies factorized scattering. This fact is the basis of the 
``bootstrap'' approach for determining $S$ matrices for integrable quantum 
field theories.

On the other hand, for integrable quantum spin chains, it is possible 
to compute the excitations' exact $S$ matrices directly from the Bethe 
Ansatz equations, without explicitly assuming factorization 
\cite{korepin} - \cite{andrei/destri}.  Therefore, it should be 
possible to explicitly demonstrate within this framework the 
factorizability of multiparticle scattering.  The purpose of this note 
is to provide such a demonstration for the antiferromagnetic 
spin-${1\over 2}$ Heisenberg spin chain
\be
H= {1\over 4}\sum_{n=1}^N \left( \vec \sigma_n \cdot  \vec
\sigma_{n+1} - 1 \right) \,, \qquad \qquad
\vec \sigma_{N+1} \equiv \vec \sigma_1 \,, \label{closed}
\ee
which is the prototypical integrable quantum spin chain.  For 
simplicity, we restrict our attention to the case of 3-body 
scattering, which is the first nontrivial case to exhibit 
factorization.  We remark, following Faddeev and Takhtajan \cite{FT1}, 
that states with an odd number of excitations appear in the sector of 
the model with $N=$ odd.\footnote{For another recent investigation 
which exploits the odd sector of this model, see Ref.  \cite{DMN}.}

The outline of this paper is as follows.  In Sec. 2 we review the 
Bethe Ansatz/string hypothesis description of multiparticle excited 
states of the Heisenberg spin chain.  We compute the three-particle 
$S$ matrix in Sec. 3; and we show that it is factorizable into a 
product of two-particle $S$ matrices in Sec. 4. Finally, in Sec. 5
we discuss the prospects of extending this analysis to the case of 
more than three particles. In the Appendix we briefly review the 
two-particle $S$ matrix.

\section{Excited States}

In this section we review the Bethe Ansatz/string hypothesis 
description of excited states of the Heisenberg Hamiltonian 
(\ref{closed}).  Adopting the string hypothesis, the Bethe ansatz 
equations lead to the following equations for the real centers 
$\lambda_\alpha^n$ of the strings (see, e.g., Refs.  \cite{FT2}, 
\cite{Nankai}):
\be
h_n ( \lambda_\alpha^n ) = J_\alpha^n \,,
\label{BAlog}
\ee
where $\alpha = 1, \cdots, M_n$ ($M_{n}$ is the number of strings of 
length $n$) and $n=1, \cdots, \infty \,.$
The so-called counting function $h_n (\lambda)$ is defined by
\be
h_n(\lambda) = {1\over 2\pi} \left\{
N q_n(\lambda) - \sum_{m=1}^\infty \sum_{\beta=1}^{M_m}  \Xi_{n m}
(\lambda - \lambda_\beta^m) \right\} \,, \label{hn}
\ee
$\Xi_{n m} (\lambda)$ is given by
\be
 \Xi_{n m} (\lambda) = (1 - \delta_{n m}) q_{|n-m|} (\lambda)
+ 2q_{|n-m|+2}(\lambda)
+ \cdots + 2q_{n+m-2}(\lambda) + q_{n+m}(\lambda) \,, \label{xi}
\ee
and $q_n (\lambda)$ is the odd monotonic-increasing function defined by
\be
q_n (\lambda) = \pi + i\log e_n(\lambda)
\,, \qquad -\pi < q_n (\lambda) \le \pi \,, \label{q}
\ee
where
\be
e_n(\lambda) = {\lambda + {in\over 2}\over \lambda - {in\over 2}} \,.
\ee 
Moreover
$\{ J_\alpha^n \}$ are integers or half-odd integers which satisfy
\be
 -J_{max}^n \le J_\alpha^n \le J_{max}^n \,, \label{range}
\ee
where $J_{max}^n$ is given by
\be
 J_{max}^n = 
{1\over 2}\left( N + M_n -1 \right) - \sum_{m=1}^\infty min (m,n)\ M_m
\,. \label{jmax}
\ee
We regard $\{J_\alpha^n \}$ as ``quantum numbers'' which parametrize the
Bethe Ansatz states. For every set $\{ J_\alpha^n \}$ in the range 
given by Eq.  (\ref{range}) (no two of which are identical), we assume 
that there is a unique solution $\{ \lambda_\alpha^n \}$ (no two of 
which are identical) of Eq.  (\ref{BAlog}).

The spin eigenvalues of the Bethe Ansatz states are given by
\be
S = S^z = {N\over 2} - \sum_{n = 1}^ \infty n M_n \,. \label{mm}
\ee

The ground state is a ``filled Fermi sea,'' with $M_1 = {N\over 2}$ and 
$M_n = 0$ for $n>1$.
The number of holes (excitations) $\nu$ in a Bethe Ansatz state is given by
\be
\nu &=& {\hbox{ number of vacancies for }} J^1_\alpha {\hbox{'s}} - 
{\hbox{ number of }} J^1_\alpha {\hbox{'s}} \non \\
&=& \left( 2 J^1_{max} + 1 \right) - M_1 \,. \label{nu}
\ee
A hole with rapidity $\lambda$ has 
energy $\varepsilon(\lambda)$ and momentum $p(\lambda)$ given by
\be 
\varepsilon(\lambda) = {\pi \over 2 \cosh \pi \lambda}  \,, \qquad 
p(\lambda) = \tan^{-1} \left(\sinh \pi \lambda \right) - {\pi\over 2} 
\,,
\label{energy/momentum} 
\ee 
respectively, and has spin ${1\over 2}$.

We shall focus on the following two classes of Bethe Ansatz states:
\begin{description}

\item[(a)]  $M_1 = {N\over 2} - {\nu \over 2}$ and $M_n = 0$ for $n>1$.
This state has $\nu$ holes and $S = S^z = {\nu \over 2}$.  

\item[(b)]  $M_1 = {N \over 2} - {\nu \over 2} - 1$, $M_2 = 1$ and $M_n = 0$ 
for $n>2$. This state has $\nu$ holes and one 2-string, with 
$S = S^z = {\nu \over 2} - 1$. This state has a multiplicity given by
\end{description}
\be
{\hbox{multiplicity}} = \left( \begin{array}{c} 
{\hbox{ number of vacancies for }} J^2_\alpha {\hbox{'s}}  \\
{\hbox{ number of }} J^2_\alpha {\hbox{'s}} \end{array} \right)
=  \left( \begin{array}{c}  2 J^2_{max} + 1  \\
            M_2 \end{array} \right) = \nu - 1 \,.
\label{degeneracy}
\ee

We define the density $\sigma(\lambda)$ of roots and holes
\be
\sigma(\lambda) ={1\over N} {d h_{1}(\lambda)\over d \lambda} \,,
\label{definesigma}
\ee
which shall play an important role in the following. Passing from the sum in 
$h_{1}(\lambda)$ to an integral, we obtain an integral equation whose solution
is given by \footnote{Since $\sigma$ depends also on the hole 
rapidities, the notation $\sigma(\lambda, \tilde\lambda_1 \,, \cdots \,, 
\tilde\lambda_\nu)$ would be more accurate. However, following the usual 
practice, we suppress the dependence on the hole rapidities.}
\be
\sigma(\lambda) =  s(\lambda) + {1\over N} r(\lambda) \,,
\label{sigma}
\ee
where 
\be
r_{(a)}(\lambda) &=& \sum_{\alpha=1}^{\nu} J(\lambda - 
\tilde\lambda_{\alpha}) \,, \label{(a)} \\ 
r_{(b)}(\lambda) &=& r_{(a)}(\lambda) - a_1(\lambda - \lambda_0) \,,
\label{(b)}
\ee  
for the states (a) and (b), respectively;
$\tilde\lambda_1 \,, \cdots \,, \tilde\lambda_\nu$ are the hole
rapidities; and $\lambda_0 \equiv \lambda_{1}^{2}$ is the position of 
the center of the 2-string, which we determine below.  Moreover, we 
use the notations
\be
a_n(\lambda) = {1 \over 2 \pi} {{d q_n(\lambda)} \over {d \lambda}} \,,
\label{sth}
\ee
\be
s(\lambda) = {1\over 2 \cosh \pi \lambda}
 = {1\over 2\pi} \int_{-\infty}^\infty d\omega\ e^{-i \omega \lambda}\
{e^{-|\omega|/2}\over 1 + e^{-|\omega|}} \,, 
\label{s}
\ee
\be
J(\lambda) =
{1\over 2\pi} \int_{-\infty}^\infty d\omega\ e^{-i \omega \lambda}\
{e^{-|\omega|}\over 1 + e^{-|\omega|}} \,.
\label{definitions} 
\ee

In order to calculate the position $\lambda_0$ of the center of the 
2-string in state (b), we observe that $h_{2}(\lambda_{0}) = J_{1}^{2}$.
Passing from the sum to the integral, we eventually obtain
\be
\sum_{\alpha =1}^{\nu} q_1(\lambda_0 - 
\tilde\lambda_{\alpha}) = 2 \pi J_1^2 \,.
\ee
By exponentiating and also recalling Eq. (\ref{jmax}), we obtain the 
desired result
\be 
\prod_{\alpha = 1}^{\nu} e_1(\lambda_0 - \tilde\lambda_{\alpha}) = 1 \,.
\label{constraint}
\ee
The last equation can be solved for the position $\lambda_0$ of the 
center of the 2-string in terms of the hole rapidities.  For example, 
for the particular case of two holes $(\nu = 2)$, we obtain the 
well-known result
\be
\lambda_0 = {1\over 2}\left( \tilde\lambda_1 + \tilde\lambda_2 
\right) \,.
\label{lambda0two}
\ee 
For $\nu = 3$, there are two solutions:
\be
\lambda_0^{\pm} =  {1\over 3} \left(
\tilde\lambda_1 + \tilde\lambda_2 + \tilde\lambda_3 
{\pm} \sqrt{A + {3\over 4}} \right) \,,  
\label{lambda0}
\ee
where 
\be
A = \tilde\lambda_1(\tilde\lambda_1 - \tilde\lambda_2) + \tilde\lambda_2
(\tilde\lambda_2 - \tilde\lambda_3) + \tilde\lambda_3(\tilde\lambda_3 
- \tilde\lambda_1) \,.
\label{A}
\ee

\section{Three-particle $S$ matrix}

We now restrict our attention to states with 3 excitations.  
Recall that the excitations have spin ${1\over 2}$, and that ${1\over 
2}\bigotimes{1\over 2}\bigotimes{1\over 2} = {1\over 
2}\bigoplus{1\over 2}\bigoplus{3\over 2}$.  The state with $S = S^z = 
{3\over 2}$ can evidently be identified as the Bethe Ansatz state (a) 
introduced in the previous section with $\nu = 3$.  Moreover, the two 
states with $S = S^z = {1 \over 2}$ are the Bethe Ansatz states (b). 
Indeed, the multiplicity factor given in Eq. (\ref{degeneracy}) for $\nu = 3$ 
is equal to 2.

Following Refs. \cite{korepin}, \cite{andrei/destri},
we define the $S$ matrix $S({\tilde\lambda_1, \tilde\lambda_2, 
\tilde\lambda_3})$ (acting in $C^2 \bigotimes C^2 \bigotimes C^2$) 
for a hole of rapidity $\tilde\lambda_1$ scattering with holes of 
rapidities $\tilde\lambda_2$ and $\tilde\lambda_3$ by the momentum
quantization condition
\be
\left(e^{i p(\tilde\lambda_1) N}\
S({\tilde\lambda_1, \tilde\lambda_2, \tilde\lambda_3}) - 1 \right) 
| \tilde\lambda_1, \tilde\lambda_2, 
\tilde\lambda_3 \rangle = 0 \,, \label{quantization} 
\ee
where the hole momentum $p(\lambda)$ is given in Eq. (\ref{energy/momentum}).

We shall compute the eigenvalues of $S$.
Let $S_{(a)}$, $S_{(b)}$ be the eigenvalues of $S$ corresponding to 
states (a), (b), respectively. For state (a),
\be
e^{i p(\tilde\lambda_1) N}\ S_{(a)} = 1 \,,
\label{eigen}
\ee 
and similarly for (b). 

We next derive a relation between $p(\tilde\lambda_1)$ and the 
quantity $r(\lambda)$ which characterizes the distribution of roots 
and holes for a state.  From Eqs.  (\ref{energy/momentum}), 
(\ref{definesigma}), (\ref{sigma}) and (\ref{s}), we obtain
\be 
{1\over 2\pi} {dp\over d\lambda} + {1\over N} r = 
{1\over N} {d h_{1}\over d\lambda} \,.
\ee
Integrating from $-\infty$ to $\tilde\lambda_{1}$ and exponentiating, 
we obtain the desired relation
\be
e^{i p(\tilde\lambda_1) N}\ 
e^{i 2\pi\int_{-\infty}^{\tilde\lambda_{1}}r(\lambda)\ d\lambda}\
e^{i 2\pi\left[h_{1}(-\infty) - \tilde J_{1}\right]}\
e^{-i N p(-\infty)} = 1 \,,
\label{desired}
\ee
where $h_{1}(\tilde\lambda_{1}) = \tilde J_{1}$.

Comparing this relation with Eq. (\ref{eigen}), we see that 
\be
S_{(a)} \sim   
e^{i 2\pi\int_{-\infty}^{\tilde\lambda_{1}}r_{(a)}(\lambda)\ 
d\lambda} \,.
\label{sim}
\ee 
Using the explicit expression for $r_{(a)}(\lambda)$ given in Eq.  
(\ref{(a)}) with $\nu = 3$, we conclude that $S_{(a)}$ is given (up 
to a rapidity-independent phase factor) by
\be
S_{(a)} = 
S_t(\tilde\lambda_1 - \tilde\lambda_2)\ S_t(\tilde\lambda_1 - \tilde\lambda_3)
\,,
\label{S(a)}
\ee
where
\be
S_t(\lambda) = 
{\Gamma \left(1 + {i\lambda\over 2} \right) \over
 \Gamma \left(1 - {i\lambda\over 2} \right)}
{\Gamma \left({1\over 2} - {i\lambda\over 2}  \right)\over
 \Gamma \left({1\over 2} + {i\lambda\over 2}  \right)} \,.
\label{triplet}
\ee
Since the state (a) has spin $S={3\over 2}$, the eigenvalue $S_{(a)}$ is 
4-fold degenerate.

Although we have determined $S_{(a)}$ only up to a 
rapidity-independent phase factor, we can compute the ratio 
$S_{(b)}/S_{(a)}$ exactly:
\be
{S_{(b)}\over S_{(a)}} =
e^{i 2\pi\int_{-\infty}^{\tilde\lambda_{1}} \left[ 
r_{(b)}(\lambda) - r_{(a)}(\lambda) \right] \ d\lambda}\
e^{i 2\pi\left[ h_{1}^{(b)}(-\infty) -  h_{1}^{(a)}(-\infty) \right]}\
e^{-i 2\pi\left( \tilde J_{1}^{(b)} - \tilde J_{1}^{(a)} \right)} 
= e_1(\tilde\lambda_1 - \lambda_0)  
\,,
\label{ratio}
\ee 
where we have used Eqs. (\ref{BAlog}), (\ref{jmax}), and (\ref{(b)}). 
(Each of the last two factors in Eq. (\ref{ratio}) equals $-1$.)
We conclude that
\be
S_{(b)}^\pm 
= e_1(\tilde\lambda_1 - \lambda_0^\pm)\ S_{(a)} \,,
\label{sct}
\ee
where $\lambda_0^\pm$ are given by Eq. (\ref{lambda0}). Since the states (b) 
have $S={1\over 2}$, the eigenvalues $S_{(b)}^\pm$ are each
2-fold degenerate.

In summary, the $S$ matrix $S({\tilde\lambda_1, \tilde\lambda_2, 
\tilde\lambda_3})$ has eigenvalues $S_{(a)}$ (4-fold degenerate) and
$S_{(b)}^\pm$ (each 2-fold degenerate).

\section{Factorization}

We now demonstrate that the three-particle $S$ matrix is factorizable.  
That is, we show that the $S$ matrix is equal (up to a unitary 
transformation and a rapidity-independent phase) to a product of $R$ matrices
\be
S({\tilde\lambda_1, \tilde\lambda_2, \tilde\lambda_3}) =
R_{12}(\tilde\lambda_1 - \tilde\lambda_2)\  
R_{13}(\tilde\lambda_1 - \tilde\lambda_3) \,,
\label{factorization}
\ee
where $R(\lambda)$ is the two-particle $S$ matrix, which can be 
written as (see Appendix)
\be
R(\lambda) = S_t(\lambda) \left(  \begin{array}{cccc}
      1 & 0 & 0 & 0 \\
      0 & {1\over 2}(1 + e_1({\lambda\over 2})) & {1\over 2}(1 - 
      e_1({\lambda\over 2})) & 0 \\
      0 & {1\over 2}(1 - e_1({\lambda\over 2})) & {1\over 2}(1 + 
      e_1({\lambda\over 2})) & 0 \\
      0 & 0 & 0 & 1 \end{array} \right) \,,
\ee
where $S_t(\lambda)$ is given by Eq. (\ref{triplet}).

We prove this by showing that the eigenvalues of the RHS of  Eq. 
(\ref{factorization}) 
coincide with those of the LHS. Indeed, by explicit calculation, we 
find that the eigenvalues of the $8 \times 8$ matrix
$R_{12}(\tilde\lambda_1 - \tilde\lambda_2)
R_{13}(\tilde\lambda_1 - \tilde\lambda_3)$ are $S_{(a)}$ (4-fold degenerate) 
and $\alpha^\pm S_{(a)}$ (each 2-fold degenerate), where
\be
\alpha^\pm  = {1\over 8}(B \pm \sqrt{B^2 - 64\gamma_1 \gamma_2})  \,,
\label{eig2}
\ee
and
\be
B = 1 + 3 \gamma_1 + 3 \gamma_2 + \gamma_1 \gamma_2 \,,
\label{B}
\ee
with
\be
\gamma_1 &=& e_1({{\tilde\lambda_1 - \tilde\lambda_2} \over 2}) \,, \non\\
\gamma_2 &=& e_1({{\tilde\lambda_1 - \tilde\lambda_3} \over 2}) 
\label{gamma} \,.
\ee 
By virtue of the algebraic identity\footnote{Similar identities have 
appeared in connection with scattering of excitations in open spin chains 
\cite{GMN}.}
\be
e_1(\tilde\lambda_1 - \lambda_0^\pm) 
= {1 \over 8} ( B \pm \sqrt{B^2 - 64 \gamma_1 \gamma_2}) \,,
\label{identity}
\ee
(where $\lambda_0^\pm$ are given by Eq. (\ref{lambda0})), we see that 
\be
S_{(b)}^\pm = \alpha^\pm S_{(a)} \,.
\ee
Therefore, the eigenvalues of $R_{12}(\tilde\lambda_1 - \tilde\lambda_2)
R_{13}(\tilde\lambda_1 - \tilde\lambda_3)$ indeed coincide with those of 
$S({\tilde\lambda_1, \tilde\lambda_2, 
\tilde\lambda_3})$, and Eq. (\ref{factorization}) is proved.

Note that by combining Eqs.  (\ref{quantization}) and 
(\ref{factorization}), we obtain
\be
\left( e^{i p(\tilde\lambda_1) N} R_{12}(\tilde\lambda_1 - \tilde\lambda_2)
R_{13}(\tilde\lambda_1 - \tilde\lambda_3) - 1 \right)
|\tilde\lambda_1, \tilde\lambda_2, \tilde\lambda_3 \rangle = 0 \,,
\label{q2}
\ee
which is Yang's formula \cite{yang}, \cite{ZZ2} for the case of 3 particles.

\section{Discussion}

We have seen that for the case of three holes ($\nu = 3$), the proof of 
factorizability of the $S$ matrix rests on an algebraic identity 
given by Eq. (\ref{identity}). We expect that it should be possible to extend 
this analysis to the case $\nu > 3$. The corresponding 
identities will presumably be more complicated. Indeed, for 
sufficiently large $\nu$, the algebraic equation (\ref{constraint}) which 
determines $\lambda_0$ will be of degree higher than 4. Nevertheless, 
the problem of determining the eigenvalues of
$R_{12}(\tilde\lambda_1 - \tilde\lambda_2)
R_{13}(\tilde\lambda_1 - \tilde\lambda_3) \cdots 
R_{1\nu}(\tilde\lambda_1 - \tilde\lambda_\nu)$ is equivalent to 
diagonalizing an inhomogeneous Heisenberg spin chain transfer 
matrix, and hence, can be solved for arbitrary $\nu$ by Bethe Ansatz.
(See, e.g., Ref. \cite{ZZ2}.)

\section{Acknowledgments}

We thank H. de Vega and F. Essler for discussions.
This work was supported in part by the National Science Foundation 
under Grant PHY-9507829.

\section{Appendix: Two-particle $S$ matrix}

The two-particle $S$ matrix $R(\lambda)$ was first computed by Faddeev 
and Takhtajan \cite{FT2}.  It can easily be obtained from 
Secs.  2 and 3 of the present paper.  Indeed, the state $S = S^{z} = 1$ 
(triplet) and the singlet state $S = S^{z} = 0$ correspond to states 
(a) and (b), respectively, with $\nu = 2$.  Let $S_t$ and $S_s$ be the 
eigenvalues of $R(\lambda)$ corresponding to the triplet and singlet 
states.  $S_{t}$ is given by Eq.  (\ref{triplet}) (see Eqs. (\ref{(a)})
and (\ref{sim})), and $S_{s}$ is given by
\be
S_s(\lambda) = e_1({\lambda \over 2})\ S_t(\lambda)
\label{gm}
\ee
(see Eqs. (\ref{lambda0two}) and (\ref{ratio})).

We organize these eigenvalues into a $4\times 4$ matrix as follows: by $SU(2)$ 
symmetry, $R(\lambda)$ has the general form
\be
R(\lambda) = \alpha I + \beta P  \,,
\label{matrix}
\ee
where $I$ is the unit matrix, and $P$ is the permutation 
matrix. The eigenvalues of this matrix are $\alpha + \beta$ (3-fold 
degenerate) and $\alpha - \beta$, which we identify as $S_{t}$ and 
$S_{s}$, respectively. It follows that
\be 
\alpha &=& {1\over 2} (S_t + S_s) \,, \non\\
\beta  &=& {1\over 2} (S_t - S_s) \,.
\label{app}
\ee

\vfill\eject

\vfill\eject

\end{document}